\documentclass{ws-ijmpd}

\newcommand{\gsimB}{\stackrel{\scriptstyle >}{\phantom{}_{\sim}}}     
\newcommand{\lsimB}{\stackrel{\scriptstyle <}{\phantom{}_{\sim}}}

\begin{document}

\markboth{D.\ Blaschke, T.\ Klaehn, and F.\ Weber} {Constraints on the
   nuclear EoS from neutron star observables}

\catchline{}{}{}{}{}

\title{CONSTRAINTS ON THE HIGH-DENSITY NUCLEAR EQUATION OF STATE FROM
  NEUTRON STAR OBSERVABLES}

\author{DAVID BLASCHKE} \address{Instytut Fizyki Teoretycznej,
  Uniwersytet Wroclawski, pl.\ M.\ Borna 9, 50-204 Wroclaw, Poland \\
Bogoliubov Laboratory for Theoretical Physics, JINR, 141980 Dubna, Russia \\
blaschke@ift.uni.wroc.pl}
\author{THOMAS KL{\"{A}}HN} \address{Physics Division, Argonne National
  Laboratory, Argonne, Illinois 60439-4843, USA \\
  thomas.klaehn@googlemail.com}
\author{FRIDOLIN WEBER} \address{Dept. Physics, San Diego State
  University, 5500 Campanile Drive, San Diego, CA 92182, USA \\
fweber@sciences.sdsu.edu}

\maketitle

\begin{history}
\received{Day Month Year}
\revised{Day Month Year}
\comby{Managing Editor}
\end{history}

\begin{abstract}
  Depending on the density reached in the cores of neutron stars, such objects
  may contain stable phases of novel matter found nowhere else in the
  Universe. This article gives a brief overview of these phases of matter and
  discusses astrophysical constraints on the high-density equation of state
  associated with ultra-dense nuclear matter.
\end{abstract}


\section{Introduction}

\begin{figure}[tb]
\begin{center}
  \includegraphics[keepaspectratio,width=0.65\textwidth,angle=0]
  {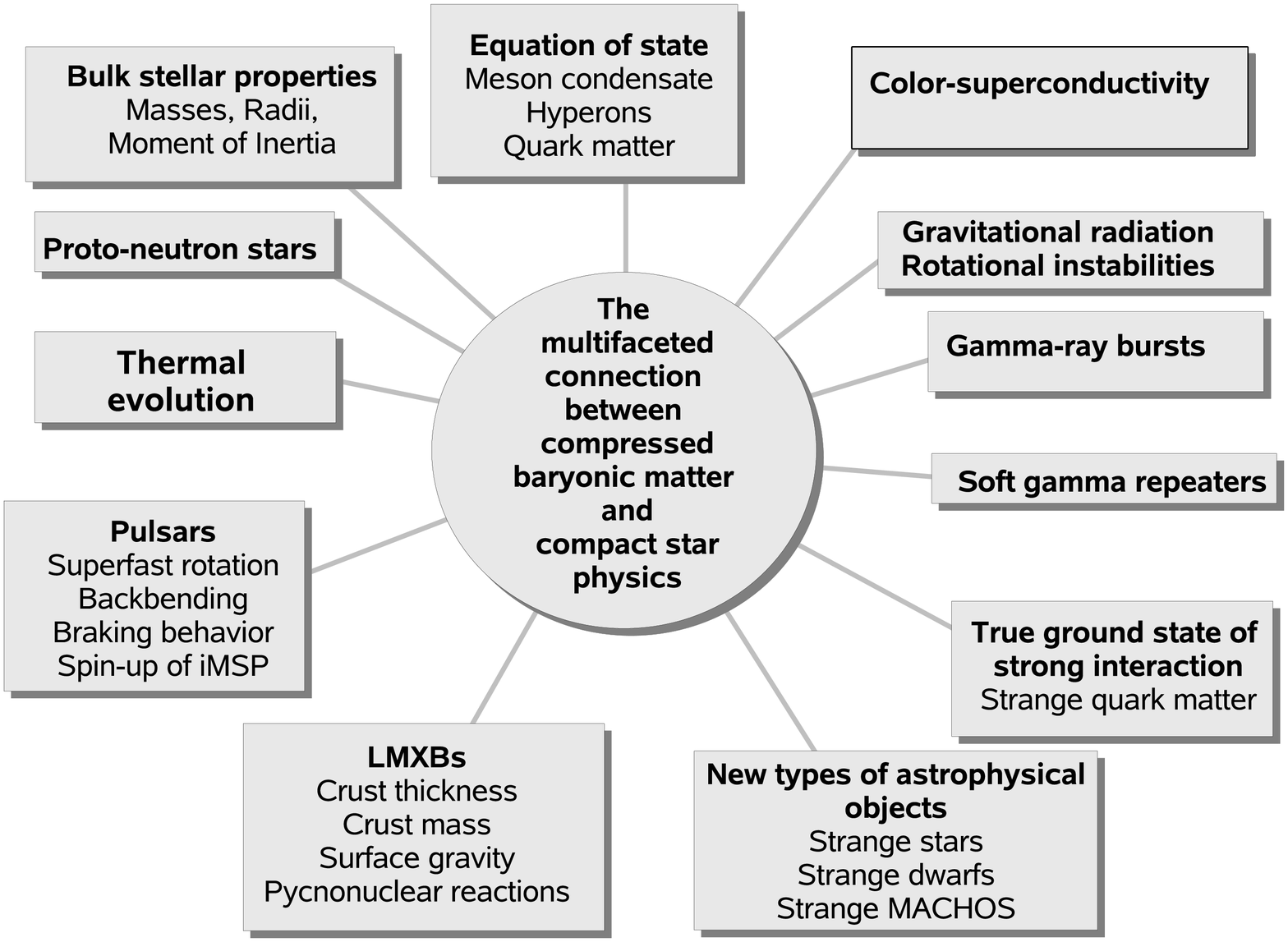}
  \caption[]{The multifaceted connection between high-density nuclear
    matter and neutron (compact) star phenomena.\cite{weber05:ppnp}}
\label{eos-blaschke-fig:multifaceted2}
\end{center}
\end{figure}

A forefront area of modern research concerns the exploration of the properties
of ultra-dense nuclear matter and the determination of the equation of state
(EoS)--the relation between pressure, temperature and density--of such matter.
Experimentally, relativistic heavy-ion collision experiments enable physicists
to cast a brief glance at hot and ultra-dense matter for times as short as
about $10^{-22}$~seconds. This is different for neutron stars, which are
observed with radio and X-ray telescopes as radio pulsars and X-ray
pulsars. The matter in the cores of such objects is compressed permanently to
densities that may be more ten times higher than the densities inside
atomic nuclei, which make neutron stars natural astrophysical laboratories
that allow for a wide range of (astro) physical studies and astrophysical
phenomena (Fig.\ \ref{eos-blaschke-fig:multifaceted2}) linked to the
properties of ultra-dense nuclear matter and its associated
EoS.\cite{glen97:book,Weber:1999qn,blaschke01:trento,%
  weber05:ppnp,Sedrakian:2006mq,page06:review,weber07:erice,Lattimer07:a,%
Schaffner07:a}
Of particular interest are neutron stars whose observed properties deviate
significantly from the norm. Examples of such neutron stars are PSR J0751+1807
whose mass is $2.1 \pm 0.2~M_\odot$,\cite{NiSp05} neutron star RX J1856.5-3754
whose radius may be $\gsimB 13$~km,\cite{Trumper:2003we} and XTE J1739-285
whose rotation period may be as small as 0.89~ms.\cite{Kaaret:2006gr} As
discussed in this paper, such neutron star data provide an excellent
opportunity to gain profound insight into the properties of nuclear matter at
most extreme conditions of density.\cite{Klahn:2006ir,Klahn:2006iw}

\section{Neutron Star Masses}\label{eos-blaschke-sec:masses}

Recent results of timing measurements for PSR B1516+02B, located in the
globular cluster M5, imply a pulsar mass of $1.96^{+0.09}_{-0.12}~{\rm
  M_\odot}$ (at 68\% probability) with a 95\% probability that the mass of
this object is above $1.68~{\rm M_\odot}$ \cite{Freire:2007sg} which deviates
considerably from the average masses of binary radio pulsars, $M_{BRP}=1.35\pm
0.04~M_{\odot}$.\cite{Thorsett:1998uc} This striking result constrains neutron
star masses to at least $1.6 ~M_{\odot}$ ($2\sigma$ confidence level), and
even to $1.9 ~M_{\odot}$ at the $1\sigma$ confidence level. The mass and
structure of spherical, non-rotating neutron stars is calculated by solving the
Tolman-Oppenheimer-Volkoff (TOV) equation,
\begin{eqnarray} 
\label{eos-blaschke-eq:TOV1} 
\frac{{\rm d}P(r)}{{\rm d}r} 
= -~\frac{ (\varepsilon(r)+ P(r))  (m(r)+ 4\pi r^3 P(r))} {r (r-2m(r))}~,
\label{eos-blaschke-eq:TOV2} 
\end{eqnarray} 
where $m(r) = 4\pi\int_0^r{\rm d}r^\prime {r^\prime}^2
\varepsilon(r^\prime)~$
is the gravitational mass inside a sphere of radius $r$.
The baryon number enclosed by this sphere is given by
\begin{eqnarray} 
\label{eos-blaschke-eq:TOV3} 
N(r)&=& 4\pi\int_0^r {\rm d} r^\prime \; {r^\prime}^2 
\left( 1 - {\frac{2m (r^\prime)}{r^\prime}} \right)^{-1/2}   
n(r^\prime) ~ ,
\end{eqnarray} 
with $n(r)$ the baryon density profile of the star. In order to solve
the TOV equation one needs to specify the stellar EoS, i.e., the
relation between pressure, $P$, and energy density, $\varepsilon$. We
apply a broad, modern collection of nuclear equations of state, which
are compiled in Table 1.\cite{Klahn:2006ir}
\begin{table}[htb]
  \tbl{Collection of nuclear equations of state studied in this
    paper. The entries are: saturation density, $n_s$; binding energy, $a_{V}$;
    incompressibility, $K$; skewness parameter, $K^{\prime}$; symmetry energy,
    $J$; symmetry energy derivative, $L$; Dirac effective mass, $m_{D}$.}
%\vspace{3mm}
{\begin{tabular}{@{}l|llrrrrr@{}} \toprule
  EoS           & $n_s$         & $a_{V}$   & $K$     & $K^{\prime}$ & $J$   & $L$   & $m_{D}$\\
                & [fm$^{-3}$]   & [MeV]     & [MeV]   & [MeV]        &[MeV]  & [MeV]  & [$m$]\\
 \hline
 NL$\rho$       & $0.1459$      & $-16.062$ & $203.3$ & $ 576.5$     &$30.8$ & $83.1$ & $0.603$ \\
 NL$\rho\delta$ & $0.1459$      & $-16.062$ & $203.3$ & $ 576.5$     &$31.0$ & $92.3$ & $0.603$ \\
 %DD             & $0.1487$      & $-16.021$ & $240.0$ & $-134.6$     &$32.0$ & $56.0$ & $0.565$ \\ 
 D${}^{3}$C     & $0.1510$      & $-15.981$ & $232.5$ & $-716.8$     &$31.9$ & $59.3$ & $0.541$ \\
% DD-F           & $0.1469$      & $-16.024$ & $223.1$ & $ 757.8$     &$31.6$ & $56.0$ & $0.556$ \\ 
 DD-F4           & $0.1469$     & $-16.028$ & $220.4$ & $ 1229.2$     &$32.7$ & $58.7$ & $0.556$ \\
 KVR            & $0.1600$      & $-15.800$ & $250.0$ & $ 528.8$     &$28.8$ & $55.8$ & $0.805$ \\
 KVOR           & $0.1600$      & $-16.000$ & $275.0$ & $ 422.8$     &$32.9$ & $73.6$ & $0.800$ \\
 DBHF           & $0.1810$      & $-16.150$ & $230.0$ & $ 507.9$     &$34.4$ & $69.4$ & $0.678$ \\
 BBG            & $0.1901$      & $-14.692$ & $221.6$ & $ -132.4$    &$36.3$ & $79.4$ & $-$        \\
 DD-RH          & $0.172$       & $-15.73$  & $249.0$ & $-$          &$34.4$ & $90.2$ & $0.686$ \\
\botrule
\end{tabular}}
\end{table}
Each one of these models was combined at sub-nuclear densities with
the Baym-Pethick-Sutherland EoS.\cite{BaPe71} The stellar radius, $R$,
is defined as that stellar location where pressure vanishes, $P(R)=0$.
The star's gravitational mass and its total baryon number are thus
given by $M=m(R)$ and $N=N(R)$, respectively.
Most of the models shown in Table~1 are derived in the framework of the
relativistic mean-field approach,\cite{Wal74,Ser86,Rei89,Rin96} allowing for
non-linear (NL) self-interactions of the $\sigma$ meson
\cite{Gaitanos:2003zg}.  For model NL$\rho$, the isovector part of the
interaction is described entirely in terms of $\rho$ meson exchange. This is
different for NL$\rho\delta$ where the isovector part of the interaction is
described in terms of both $\rho$ and $\delta$-meson exchange. The latter is
generally neglected in RMF models.\cite{Liu02} RMF models with density
dependent input parameters (coupling constants and masses) are represented in
Table 1 by four different models from two classes, where in the first one
density dependent meson couplings are modeled such that several properties of
finite nuclei (binding energies, charge and diffraction radii, surface
thicknesses, neutron skin in ${}^{208}$Pb, spin-orbit splittings) can be
fitted.\cite{Typel:2005ba} D${}^{3}$C additionally contains a derivative
coupling which leads to momentum-dependent nucleon self-energies, and DD-F4 is
modeled such that the flow constraint from heavy-ion collisions is
fulfilled.\cite{Danielewicz:2002pu} The second class of these models is
motivated by the Brown-Rho scaling assumption\cite{Brown:1991kk} that not only
the nucleon mass but also the meson masses should decrease with increasing
density.  In the KVR and KVOR models\cite{Kolomeitsev:2004ff} these
dependences are related to a nonlinear scaling function of the $\sigma$-meson
field such that the EoS of symmetric nuclear matter and pure neutron matter
below four times the saturation density coincide with those of the
Urbana-Argonne group.\cite{APR} In this way the latter approach builds a
bridge
\begin{figure}[tb]
\begin{center} 
  \includegraphics[keepaspectratio,height=0.70\textwidth,
  angle=-90]{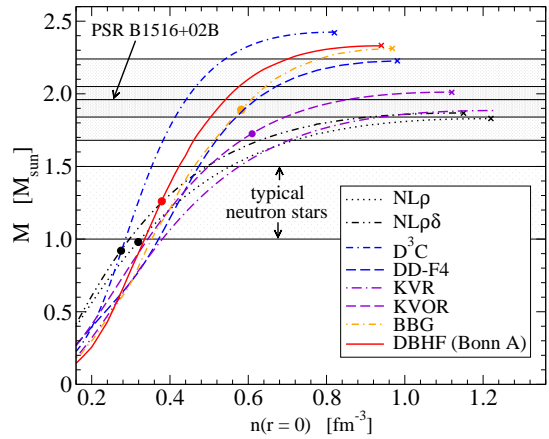}
  \caption[a]{Mass versus central density of neutron stars, computed from Eq.\
    (\ref{eos-blaschke-eq:TOV1}) for EoS shown in Table~1.  Crosses refer to
    the maximum-mass star of each sequence, filled dots mark the critical
    masses and central densities beyond which the direct Urca (DU) cooling
    process becomes possible.}
  \label{eos-blaschke-fig:M_n} 
\end{center}
\end{figure} 
between the phenomenological RMF models and a microscopic EoS built on
realistic nucleon-nucleon forces. The RMF models are contrasted with several
variational models for the EoS such as APR\cite{APR},
WFF\cite{Wiringa:1988tp}, FPS\cite{Friedman:1981qw}, a relativistic
Dirac-\-Brueckner-\-Hartree-\-Fock (DBHF) model\cite{DaFuFae04}, and a
nonrelativistic Brueckner-\-Bethe-\-Goldstone (BBG) model.\cite{Baldo:1999rq}
As shown in Fig.\ \ref{eos-blaschke-fig:M_n}, none of these values falls below
the $2\sigma$ mass limit of $1.68 ~M_\odot$ for PSR B1516+02B, and even at the
$1\sigma$ mass limit of $1.84 ~M_\odot$ the softest EoS NL$\rho$ and
NL$\rho\delta$ cannot be excluded.  We point out that if a pulsar with a mass
exceeding $1.8-1.9~M_\odot$ at the 2$\sigma$ or even 3$\sigma$ level would be
observed in the future, this would impose severe constraints on the stiffness
of the nuclear EoS. For the set of EoS tested here, only the stiffest 
models, i.e. D$^3$C, DD-F4, BBG, and DBHF would remain viable candidates.

\section{Gravitational Mass--Baryon Number Relation}
 
It has been suggested that pulsar~B in the double pulsar system J0737--3039
may serve to test models proposed for the EoS of superdense nuclear
matter.\cite{Pod05} One of the interesting characteristics of this system is
that the mass of pulsar~B is merely $1.249 \pm 0.001
~M_\odot$.\cite{Kramer:2005ez,Faulkner05:a} Such a low mass could be an
indication that pulsar~B did not form in a type-II supernova, triggered by a
collapsing iron core, but in a type-I supernova of an ONeMg white dwarf driven
hydrostatically unstable by electron captures onto Mg and Ne.\cite{Pod05} The
well-established critical density at which the collapse of such stars sets in
is $4.5\times 10^9~{\rm g}/{\rm cm}^3$.  Assuming that the loss of matter
during the formation of the neutron star is negligible, a predicted baryon
mass for the neutron star of $M_N = 1.366 - 1.375 ~M_\odot$ was derived in
Ref.\ \refcite{Pod05}. This theoretically inferred baryon number range
together with the star's observed gravitational mass of $M = 1.249 \pm
0.001~M_\odot$ may represent a most valuable constraint on the EoS, provided
the above key assumption for the formation mechanism of the pulsar B is
correct.\cite{Pod05,B-K:2007} If so, any
\begin{figure}[tb] 
\begin{center}
  \includegraphics[keepaspectratio,height=0.70\textwidth,
  angle=-90]{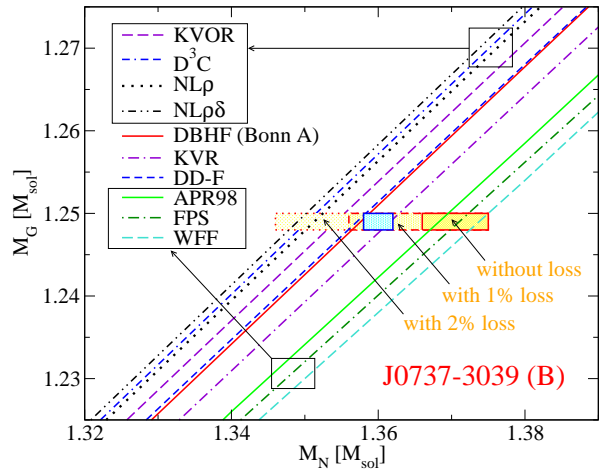}
  \caption[a]{Gravitational mass--baryon mass relationship\cite{Pod05} for PSR
    J0737-3039 (B) compared to a set of relativistic EoS
    and four nonrelativistic EoS (BBG, APR, FPS and WFF).\cite{Klahn:2006ir}}
\label{eos-blaschke-fig:M_N}
\end{center}
\end{figure} 
viable EoS proposed for neutron star matter must predict a baryon number in
the range $1.366 \lsimB M_N \lsimB 1.375 ~M_\odot$ for a neutron star whose
gravitational mass is $M=1.249 \pm 0.001~M_\odot$. We may contrast this result
with that of an independent calculation,\cite{Kitaura:2005bt} where for
pulsar~B a baryon mass of $M_N=1.360\pm 0.002~M_\odot$ has been obtained.  The
authors of Ref.\ \refcite{Pod05} discussed caveats such as baryon loss and
variations of the critical mass due to carbon flashes during the collapse.
The effect of 1\% and 2\% mass loss on the usefulness of this constraint to
exclude model EoS is shown in Fig.~\ref{eos-blaschke-fig:M_N}.

\section{Mass-Radius  Constraints from Neutron Stars in
 LMXBs}
\label{eos-blaschke-subsec:mr_qpo} 

\begin{figure}[tb] 
\centering 
\includegraphics[keepaspectratio,height=0.49\textwidth,
angle=-90]{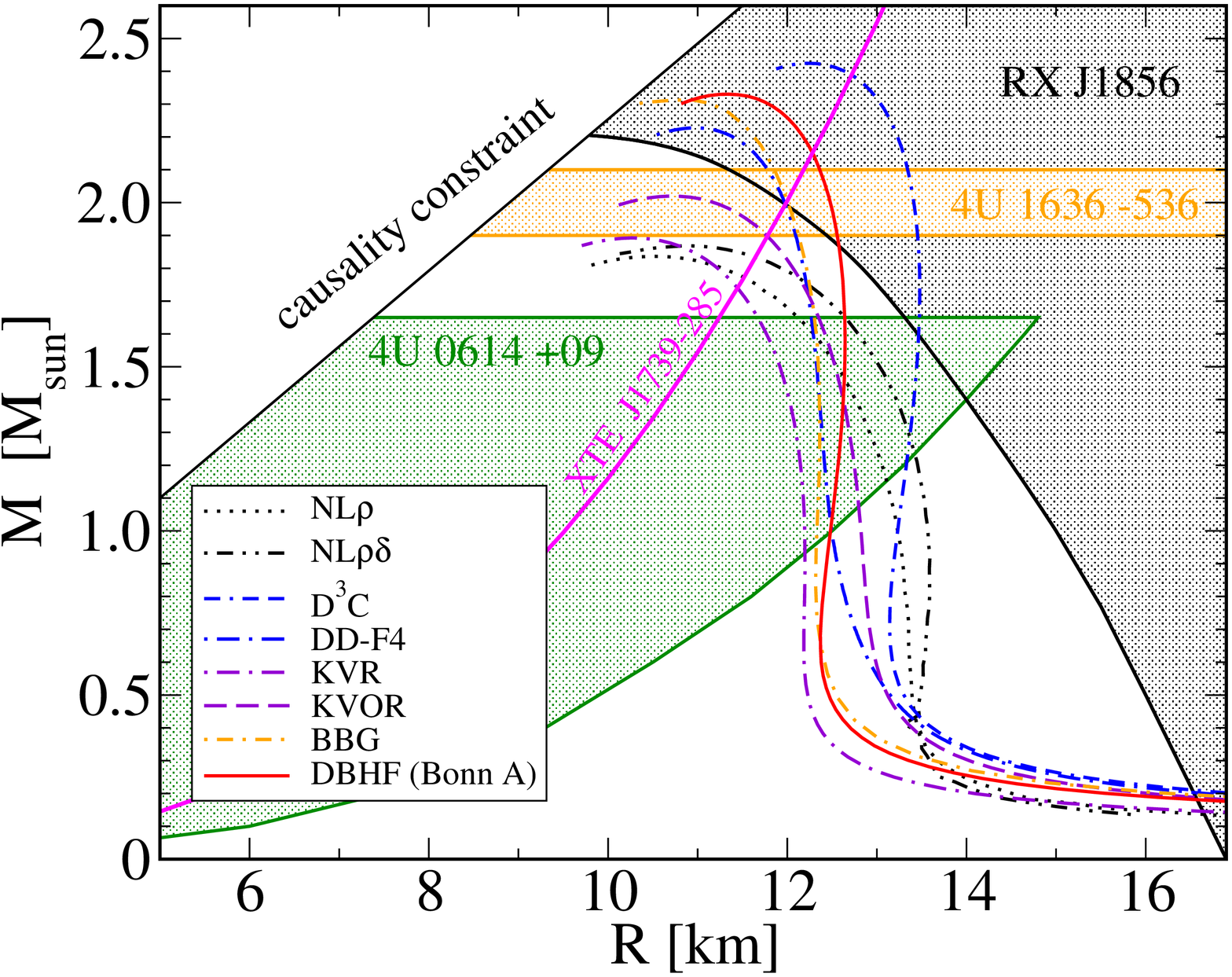} %&
\hfill 
\includegraphics[keepaspectratio,height=0.49\textwidth,
angle=-90]{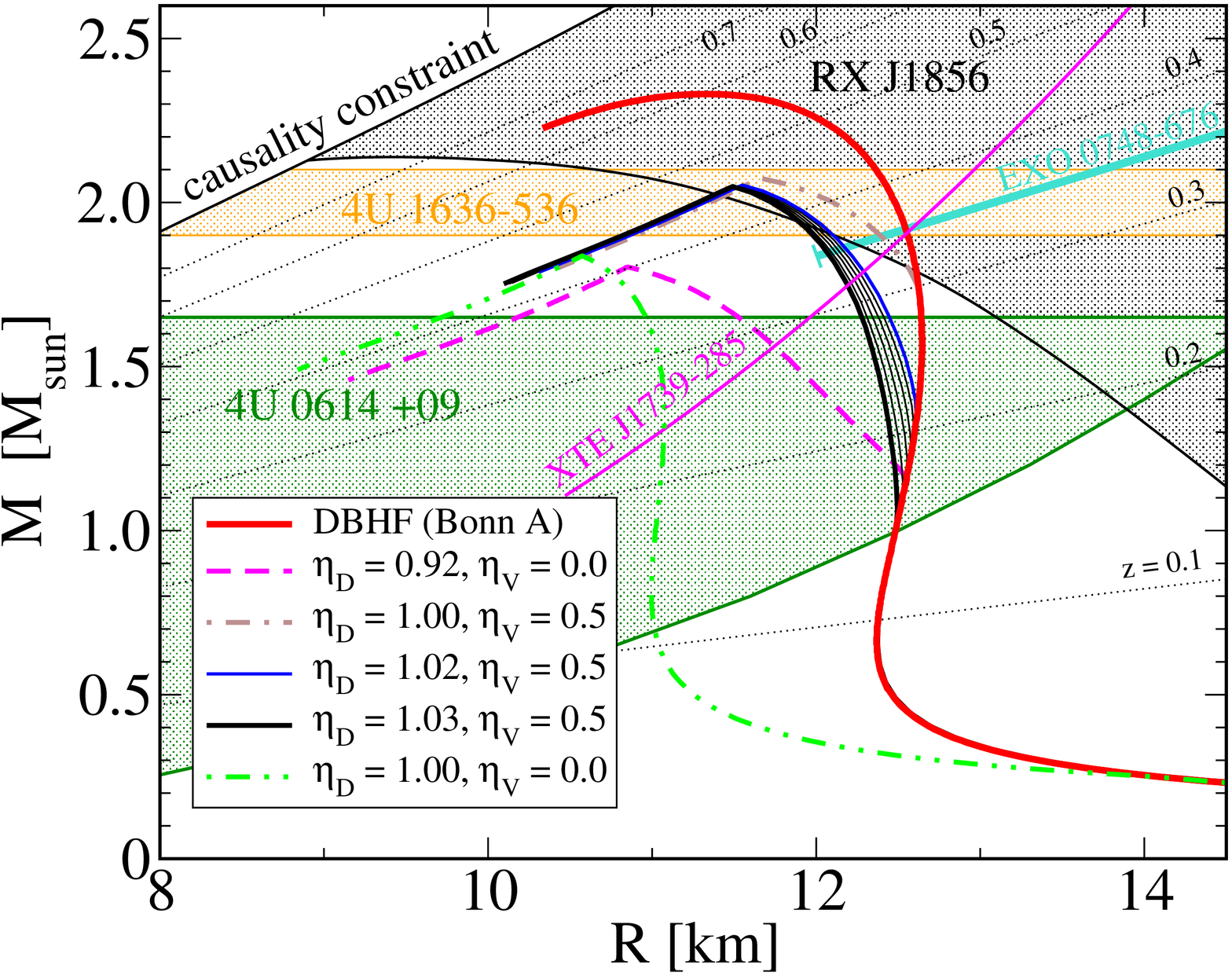}
\caption[a]{Left panel: Mass-radius constraints from thermal radiation of the
  isolated neutron star RX J1856 (grey hatched region) and from QPOs in the
  LMXBs 4U 0614+09 (wedge-like green-hatched area) and 4U 1636-536 (orange
  hatched region).  For 4U 1636-536 a mass of $2.0\pm 0.1~M_\odot$ is obtained
  so that the weak QPO constraint would exclude the NL$\rho$ and
  NL$\rho\delta$ EoS, whereas the strong one renders only the stiffest EoS
  D$^3$C, DD-F4, BBG and DBHF viable. XTE J1739-285 would favor a soft EoS,
  which is controversial, however (see text). Right panel: Same as left panel,
  but for hybrid stars with a DBHF hadronic shell and a 2SC color
  superconducting NJL quark matter core computed for different coupling
  strengths, $\eta_D$ and $\eta_V$.  The dotted lines display surface
  redshifts, $z$, ranging from $0.1, \ldots, 0.7$.}
\label{eos-blaschke-fig:M-R} 
\end{figure}

Aside from neutron star masses, kilohertz quasi-periodic brightness
oscillations (QPOs) seen from more than 25 neutron star X-ray binaries (LXMBs)
can be used to put additional constraints on the high-density EoS. A pair of
such QPOs is often seen from these systems.\cite{vdK:2000} In all currently
viable models for these QPOs, the higher QPO frequency is close to the orbital
frequency at some special radius. For such a QPO to last the required many
cycles (up to $\sim 100$ in some sources), the orbit must be outside the
star. According to general relativity theory the orbit must also be outside
the innermost stable circular orbit (ISCO).  Gas or particles inside the ISCO
would spiral rapidly into the star, preventing the production of sharp QPOs.
This implies\cite{Miller:2003wa,Miller:1996vj} that the observation of a
source whose maximum QPO frequency is $\nu_{\rm max}$ limits the stellar mass
and radius to
\begin{equation} 
\begin{array}{rl} 
  M < 2.2~M_\odot {{1000~{\rm Hz}} \over {\nu_{\rm max}}} 
(1+0.75j) ~ , \quad
  R < 19.5~{\rm km}
{{1000~{\rm Hz}} \over {\nu_{\rm max}}}
(1+0.2j) ~ .
\label{eq:qpo}
\end{array} 
\end{equation} 
The quantity $j\equiv cJ/GM^2$ (with $J$ the stellar angular momentum) is the
dimensionless spin parameter, which is typically in the range between 0.10 and
0.2 for these systems. Equation (\ref{eq:qpo}) implies that for given observed
value $\nu_{\rm max}$ the mass and radius of that source must be inside a
wedge-shaped area, as shown in Fig.\ \ref{eos-blaschke-fig:M-R}. Since wedge
becomes smaller for higher $\nu_{\rm max}$, the highest frequency ever
observed, 1330~Hz for 4U~0614+091\cite{vanS:2000}, places the strongest
constraint on the EoS.  As can be seen from Fig.\ \ref{eos-blaschke-fig:M-R},
the current QPO constraints do not rule out any of the EoS considered
here. However, because higher frequencies imply smaller wedges, the future
observation of a QPO with a frequency in the range of $\sim 1500-1600$~Hz
would rule out the stiffest of our EoS.
 
If there is evidence for a particular source that a given frequency is
close to the orbital frequency at the ISCO, then the mass is known to
a good accuracy, with uncertainties arising from the spin parameter.
This was first claimed for 4U~1820--30,\cite{Zhang:1998} but
complexities in the source phenomenology have made this controversial.
More recently, careful analysis of Rossi X-ray Timing Explorer data
for 4U~1636--536 and other sources\cite{Barret:2005wd} has suggested
that sharp and reproducible changes in QPO properties are related to
the ISCO.  If so, this implies that several neutron stars in low-mass
X-ray binaries should have gravitational masses between $1.9\,M_\odot$
and possibly $2.1\,M_\odot$.\cite{Barret:2005wd} In Fig.\
\ref{eos-blaschke-fig:M-R} we show the estimated mass of $2.0\pm
0.1\,M_\odot$ for 4U~1636--536.

Recently, mass-radius constraints have been reported for the accreting
compact stars XTE J1739-285\cite{Lavagetto:2006ew} and SAX
J1808.4-3658\cite{Leahy:2007fb} (SAX J1808 for short) which are based
on the identification of the burst oscillation frequency with the spin
frequency of the compact star.  
%As can be seen from Fig.\ \ref{eos-blaschke-fig:M-R}, 
It is almost impossible to fulfill the
constraints from RX J1856 (and other high-mass candidates) and SAX
J1808 (which favors a soft EoS) simultaneously so that the status of
SAX J1808 is currently controversial. It is likely that the small
radius estimate of Leahy et al.\cite{Leahy:2007fb} is a consequence of
the underestimation of higher harmonics when only timing data are
analyzed and not also the energy spectra.\cite{Miller:2007} Neutron
star XTE J1739-285 and its possible implications for the EoS are
discussed in more detail below.

\section{Mass-Radius Relation Constraint from RX J1856.5-3754} 
\label{eos-blaschke-subsec:trumper} 

The nearby isolated neutron star RX J1856.5-3754 (RX J1856 for short) belongs
to a group of seven objects which show a purely thermal spectrum in X-rays and
in optical-UV.  This allows the determination of $R_\infty/d$, the ratio of
the photospheric radius $R_\infty$ to the distance $d$ of the object, if the
radiative properties of its photosphere are known.  RX J1856 is the only
object of this group which has a measured distance obtained by Hubble Space
Telescope (HST) astrometry.  After the distance of 117~pc \cite{Walter:2002uq}
became known several groups pointed out that the blackbody radius of this star
is as large as 15 to 17~km.  Although both the X-ray and the optical-UV
spectra are extremely well represented by blackbody functions they require
different emission areas, a smaller hot spot and a larger cooler region.  The
overall spectrum could also be fitted by blackbody emission from a surface
showing a continuous temperature distribution between a hot pole and a cool
equator, as expected for a magnetized neutron star. The resulting blackbody
radii are 17~km (two blackbodies) and 16.8~km (continuous temperature
distribution)\cite{Trumper:2003we}. In this paper we adopt the result of the
continuous temperature fit, $R_\infty=16.8$ km.  More recent HST observations
of RX J1856 indicated larger distances of up to 178~pc. A distance of around
140~pc for RX J1856 is considered a conservative lower
limit.\cite{Kaplan:2006} For a distance of 140~pc the corresponding radius is
17~km.\cite{Ho:2006} Although some questions--in particular that of the
distance--are not yet finally settled, the recent data support an unusually
large radius for RX J1856.5-3754.

\section{Surface Redshift} 

An additional test to the mass-radius relationship is provided by
measurements of the gravitational redshift of line emissions from the
surfaces of neutron stars. The gravitational redshift, $z$ is given in
terms of the star's gravitational mass and radius according to
\begin{equation} 
\label{eos-blaschke-redshift} 
z = 1 / \sqrt{1-2M/R} - 1~. 
\end{equation}  
In the right panel of Fig.\ \ref{eos-blaschke-fig:moi}, the dependence of $z$
on star mass is shown for two representative EoS. The disputed measurement of
$z=0.35$ for EXO 0748-676\cite{Cottam:2002,Ozel:2006km} allows for both an
unconfined hadronic core composition as well as a deconfined quark-core
composition.\cite{Alford:2006vz} A measurement of $z\ge 0.5$ could not be
reconciled with the hybrid star models suggested here, while the hadronic
model would not be invalidated by redshift measurements up to $z=0.6$.

\section{Moment of Inertia} 

\begin{figure}[tb]     
\centering 
%\begin{tabular}{cc} 
\includegraphics[keepaspectratio,height=0.49\textwidth,
angle=-90]{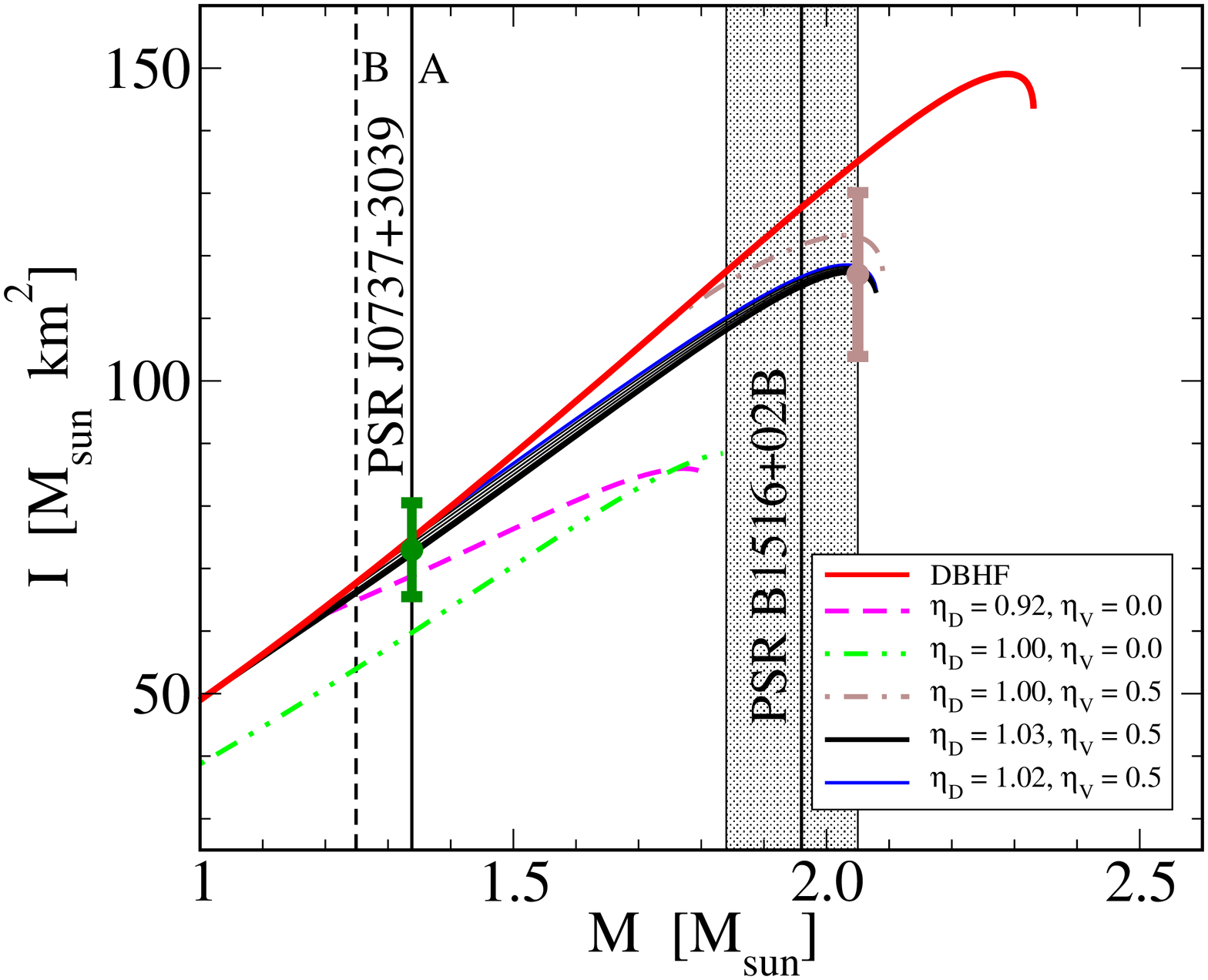} %&
\hfill    
\includegraphics[keepaspectratio,height=0.49\textwidth,
angle=-90]{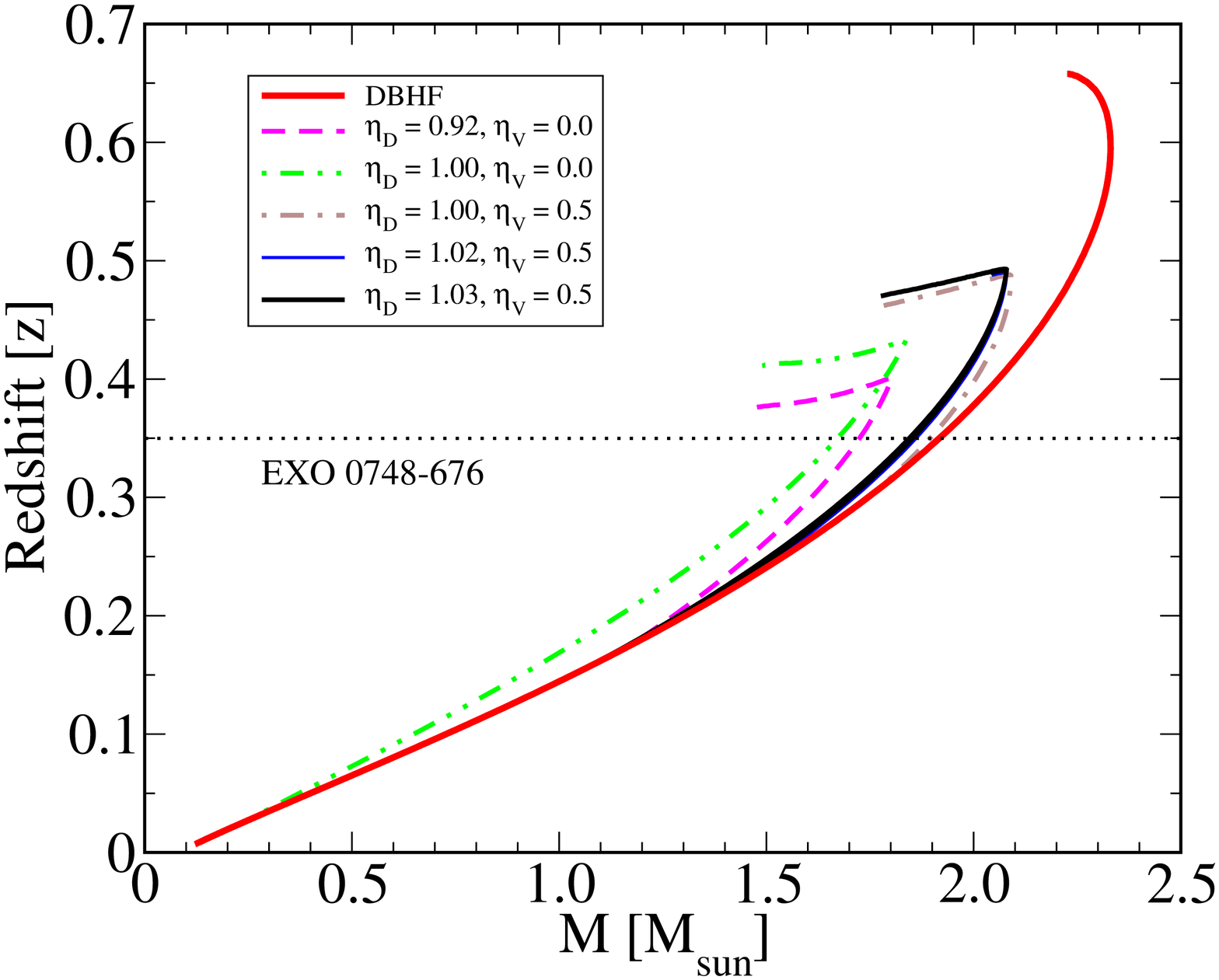}
%\end{tabular}     
\caption[a]{Left panel: Moment of inertia versus mass.  The highlighted mass
  regions correspond to the double pulsar J0737+3039 A+B and the pulsar
  B1516+02BA.  Right panel: Surface redshift versus gravitational mass for a
  purely hadronic star (DBHF) as well as hybrid stars with 2SC color
  superconducting NJL quark matter cores. The dotted line labeled EXO 0748-676
  displays the disputed redshift $z=0.35$.\cite{Cottam:2002,Ozel:2006km}}
\label{eos-blaschke-fig:moi}
\end{figure}

The observed data of the relativistic double pulsar PSR J0737+3039 did
allow a determination of the moment of inertia
(MoI)\cite{glen97:book,Weber:1999qn} of that star, which may be used to put
further constraints on the EoS of neutron
stars.\cite{Bejger:2005jy,Lattimer:2004nj} Results for the MoI
computed for the EoS of this paper are shown in Fig.~
\ref{eos-blaschke-fig:moi}.  Due to the fact that the mass 1.338
$M_\odot$ of PSR J0737+3039 A is in the vicinity of the suggested
critical mass region, the quark matter core is small and the expected
MoI of the hybrid star will be practically indistinguishable from that of a
pure hadronic one. The situation would improve if the MoI could be measured
for more massive objects, because the difference in the MoI of both
alternative models for masses as high as $2~M_\odot$ could reach the $10\%$
accuracy level.

\section{Rotational Frequencies} 

An absolute upper limit on the spin frequency of a pulsar is given by
the mass shedding limit, at which the velocity of the stellar surface
equals that of an orbiting particle suspended just above the surface.
For a rigid Newtonian sphere this frequency is given by the Keplerian
frequency,\cite{Lattimer:2004pg}
\begin{equation} 
  \label{eos-blaschke-nu_K} 
  \nu_K=(2\pi)^{-1}\sqrt{G M/R^3}=1833 ~(M/M_\odot)^{1/2}(R/10~{\rm km})^{-3/2} ~
  {\rm Hz}~. 
\end{equation} 
This formula was found to describe the mass shedding points for a sample of
neutron star EoS extremely well.\cite{Bejger:2006hn} However, both deformation
and general relativistic effects are very important so that Eq.\
(\ref{eos-blaschke-nu_K}) needs to be modified.  It has been
found\cite{Lattimer:2004pg} that with a coefficient of 1045 Hz, Eq.\
(\ref{eos-blaschke-nu_K}) approximately describes the maximum rotation rate
for a star of mass $M$ and non-rotating radius $R$, independently of the EoS.
The observation of rapidly rotating pulsars can therefore constrain the
compactness and might eventually lead to the elimination of EoS that are too
stiff, since the latter lead to stars that are too big. If the recent
discovery of burst oscillations with a frequency of 1122 Hz in the X-ray
binary XTE J1739-285\cite{Kaaret:2006gr} and their identification with the
star's spin frequency turns out to be correct, this object would spin at a
rotational period of 0.89~ms, rendering it the first sub-millisecond pulsar
ever observed, imposing severe constraints on the EoS.  The boundary for the
mass-radius relationship reads in this case\cite{Lavagetto:2006ew}
\begin{equation} 
\label{eos-blaschke-XTE1739} 
R<9.52~(M/M_\odot)^{1/3}~{\rm km}~. 
\end{equation} 
As can be seen from the left panel in Fig.\ \ref{eos-blaschke-fig:M-R}, this
constraint would rule out all neutron stars of masses $M<1.75~M_\odot$. Only
neutron stars computed for the stiffest EoS of our collection (D$^3$C, DD-F4,
BBG and DBHF) remain viable.
In concluding this section, we mention that an important test of the transport
properties of the matter inside sub-millisecond pulsars arises from the r-mode
instability.\cite{Madsen:1999ci} This constraint on superconducting quark
matter in neutron stars has recently been discussed for the putative
sub-millisecond pulsar XTE J1739-285,\cite{drago07:a} where it was argued that
this object must be either a strange star or a quark-hadron hybrid star.
\begin{figure}[tb]
\centering 
\includegraphics[keepaspectratio,height=0.9\textwidth,angle=-90]
{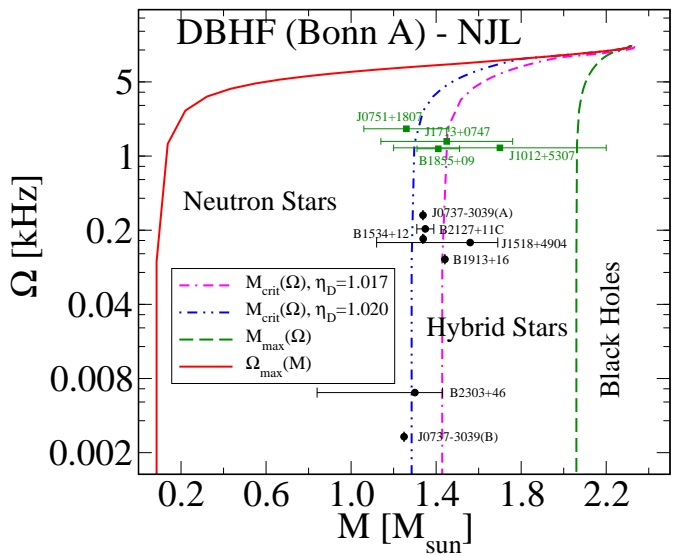}
\caption[a]{Rotational frequency, $\Omega$, versus mass, $M$, of neutron
  stars. Stable stars are located in the region bordered by the maximum
  rotation frequency $\Omega_{\rm max}(M)$, where mass-shedding from the
  equator occurs, and the maximum mass $M_{\rm max}(\Omega)$ at which
  gravitational instability against collapse to a black hole sets in. The
  dash-dotted (dash-double-dotted) line marks the onset of the generation of a
  quark matter core in the star's interior for a diquark coupling parameter
  $\eta_D=1.020$ ($\eta_D=1.017$), see Fig.\ \ref{eos-blaschke-fig:M-R} for
  the corresponding mass-radius relations of the non-rotating stellar
  sequences.  The data with error bars correspond to binary radio pulsars
  (black dots) and neutron stars with a white dwarf binary (green squares).
  The correlation of the distribution of objects with the line for a
  deconfinement phase transition could be interpreted as a waiting point
  phenomenon during the accretion evolution of the compact stars (mass
  clustering).\cite{Poghosyan:2000mr} }
\label{eos-blaschke-fig:mass-clus}
\end{figure}
In passing we mention that rotational instabilities in rotating stars, known
as gravitational radiation driven instabilities, set a more stringent limit on
rapid stellar rotation than mass shedding from the
equator.\cite{glen97:book,Weber:1999qn} These instabilities originate from
counter-rotating surface vibrational modes which at sufficiently high
rotational star frequencies are dragged forward. In this case gravitational
radiation, which inevitably accompanies the aspherical transport of matter,
does not damp the instability modes but rather drives them. Viscosity plays
the important role of damping these instabilities at a sufficiently reduced
rotational frequency such that the viscous damping rate and power in gravity
waves are comparable.  The most critical instability modes that are driven
unstable by gravitational radiation are the $f$-modes and the recently
discovered $r$-modes.\cite{andersson01:a} The latter may severely constrain
the composition of compact stars that would rotate at sub-millisecond
periods.\cite{drago07:a}

\section{Mass Clustering}

Compact stars in binary systems in general undergo during their evolution a
stage with disc accretion leading to both, spin-up and mass increase.  Initial
indications for a spin frequency clustering in low-mass X-ray binary systems,
reported by measurements with the Rossi-XTE, have lead to the suggestion to
interprete such a correlation as a waiting-point phenomenon where star
configurations cross the border between pure neutron stars and hybrid stars in
the spin frequency--mass plane.\cite{Glendenning:2000zz} A systematic analysis
of the critical line for a deconfinement phase transition in the phase diagram
for accreting compact stars\cite{Poghosyan:2000mr} has revealed that the
suggested population clustering due to the phase transition shall rather lead
to a mass clustering effect.  For strange stars, however, such an effect shall
be absent.\cite{Blaschke:2001th} For generic polytropic forms of the EoS of
quark and hadronic matter, the relationship between softness or hardness of
the EoS and the structure of this phase diagram has been demonstrated in
Ref. \refcite{Grigorian:2002ih}. For the specific example of the DBHF hadronic
EoS and a color superconducting stiff quark matter EoS, we show the phase
diagram of rotating compact stars in Fig.\ \ref{eos-blaschke-fig:mass-clus}.
\begin{figure}[tb]
\centering 
\includegraphics[keepaspectratio,height=0.49\textwidth,
angle=-90]{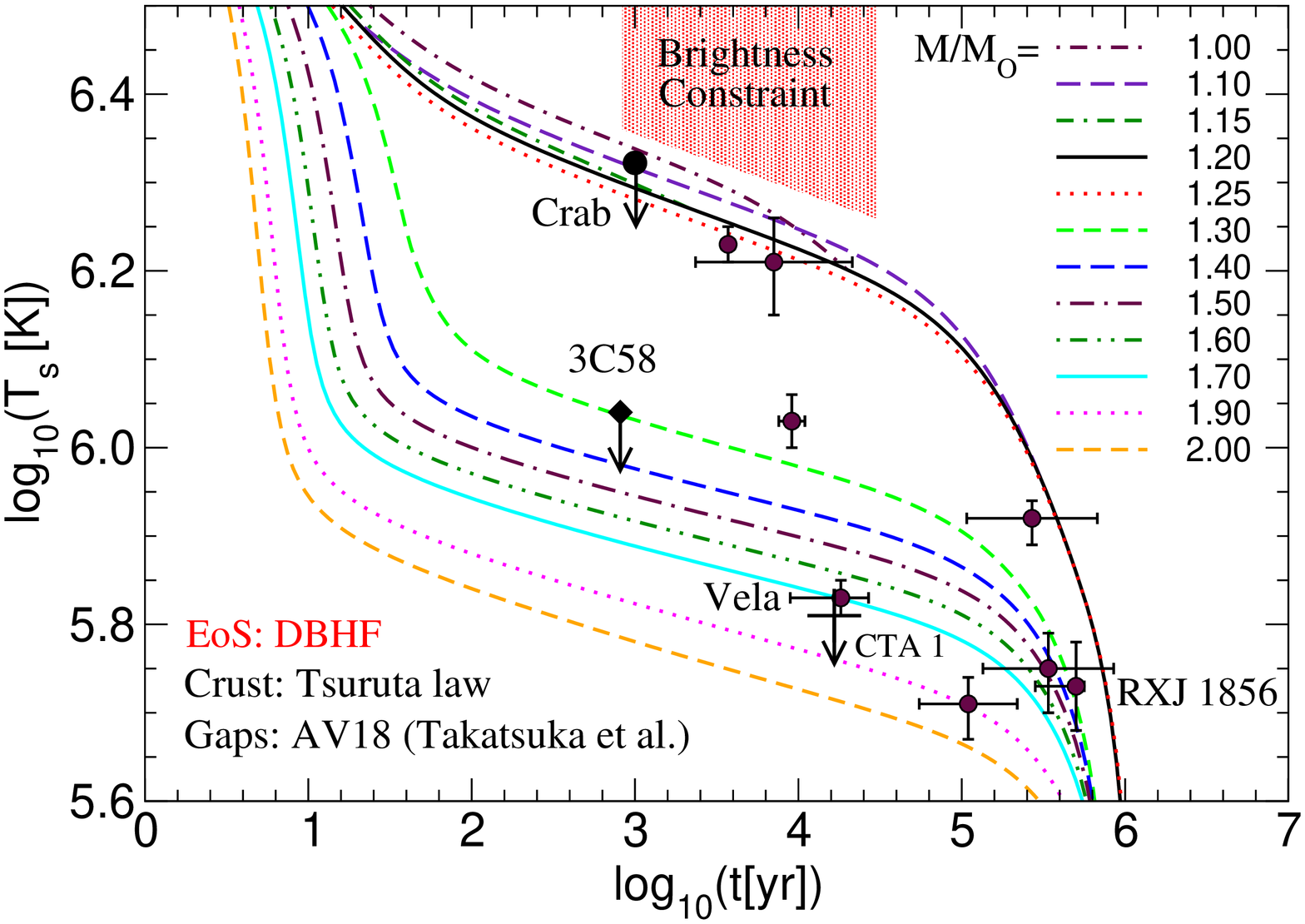} %&
\hfill 
\includegraphics[keepaspectratio,height=0.49\textwidth,
angle=-90]{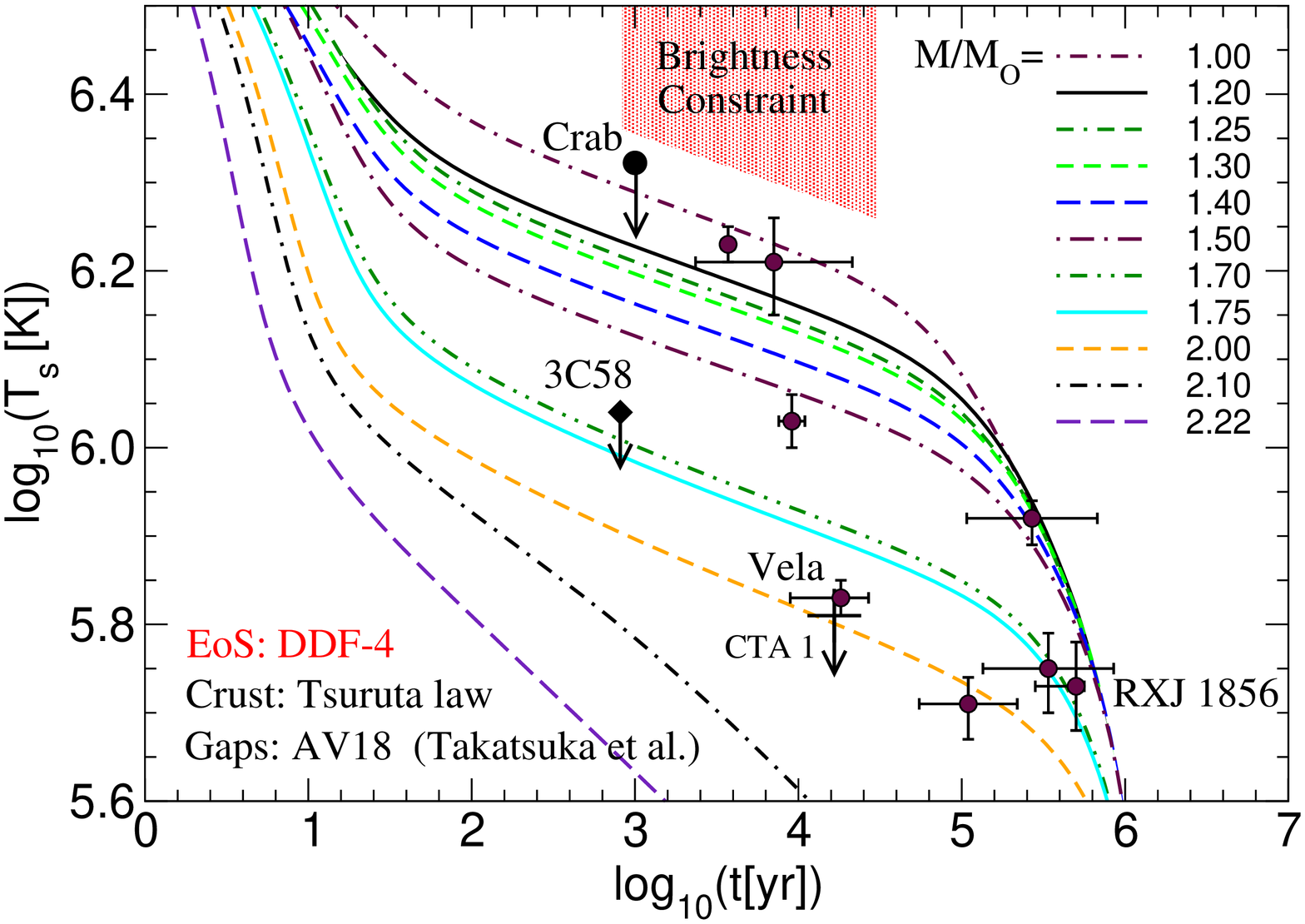}
\caption[a]{Hadronic star cooling curves for DBHF (left) and DDF-4 (right)
  model EoS.  Different lines
  correspond to compact star mass values indicated in the legend (in units of
  $M_\odot$), data points with error bars are taken from
  Ref. \cite{Page:2004fy}.  The hatched trapeze-like region represents the
  brightness constraint (BC).\cite{Grigorian:2005fd}
  The onset of the DU process entails a high sensitivity of TA curves to small
  mass variations known as DU problem\cite{Blaschke:2006gd}.
  \label{eos-blaschke-fig:hc1}}
\end{figure}
For curiosity we show also the masses and spin frequencis for compact stars in
binary radio pulsars and neutron star--white dwarf binaries and observe an
interesting correlation of the distribution of these objects with the critical
deconfinement phase transition line for diquark coupling $\eta_D=1.02$.  There
might be other reasons for the mass clustering of spinning
pulsars,\cite{Heuvel:2007zp} but the suggestion to relate it to a phase
transition in the interior can not be excluded. If true, it would be a strong
constraint for the hybrid EoS and suggest a critical density for deconfinement
in compact stars at about 0.4~fm$^{-3}$ (i.e., $2.5~n_s$).  We want to point
out that the non-zero quark chemical potential ($\mu\ne0$) domain is a rather
poorly understood region of the QCD phase diagram.  Fundamental approaches,
like solving the in-medium QCD Schwinger-Dyson equations in a specific QCD
model \cite{Roberts:2000aa,Nickel:2006vf,Chen:2008zr} to obtain a quark matter
EoS are demanding.  This well justifies the description of a quark matter
phase within three-flavor NJL-type models, giving access even to diquark
pairing channels at reasonable
expense.\cite{Blaschke:2005uj,Ruster:2005jc,Abuki:2005ms,Warringa:2005jh}

\section{Stellar  Cooling} 
\label{eos-blaschke-sec:cool} 

Neutron stars cool for the first $10^5$ years through neutrino emission from
their cores, which is followed at later times by photon emission. We simulate
this behavior by using the cooling code developed in Refs.\
\refcite{Blaschke:2004vq,Blaschke:2000dy}. The cooling calculations are
performed for different hadronic EoS, different assumptions about
superconductivity, and different stellar crust models. Moreover, we try to use
consistent microscopic inputs. (For systematic field-theoretical approaches to
the neutrino cooling problem in neutron stars see, for instance, Ref.\
\refcite{Voskresensky:2001fd,Sedrakian:2006mq,Sedrakian:2007dj,%
Voskresensky07:a}.)
The main neutrino cooling processes in hadronic matter are the direct Urca
(DU), the medium modified Urca (MMU) and the pair breaking and formation
(PBF). For quark matter, the main cooling processes are the quark direct Urca
(QDU), quark modified Urca (QMU), quark bremsstrahlung (QB) and quark pair
formation and breaking (QPFB)\cite{Jaikumar:2001hq}. Also the electron
bremsstrahlung (EB), and the massive gluon-photon decay\cite{Blaschke:1999qx}
are included.
  
The $^1S_0$ neutron and proton gaps in the hadronic shell are taken according
to the calculations in Ref.\ \refcite{Takatsuka:2004zq} corresponding to the
thick lines in Fig.\ 5 of Ref.\ \refcite{Blaschke:2004vq}. However, the
$^3P_2$ gap is suppressed by a factor 10 compared to the BCS model calculation
in Ref.\ \refcite{Takatsuka:2004zq}, consistent with arguments from a
renormalization group treatment of nuclear pairing.\cite{Schwenk:2003bc}
Without such a suppression of the $^3P_2$ gap the hadronic cooling scenario
would not fulfill the TA constraint.\cite{Grigorian:2005fn}

\begin{figure}[tb]  
\centering 
\includegraphics[keepaspectratio,height=0.49\textwidth,
angle=-90]{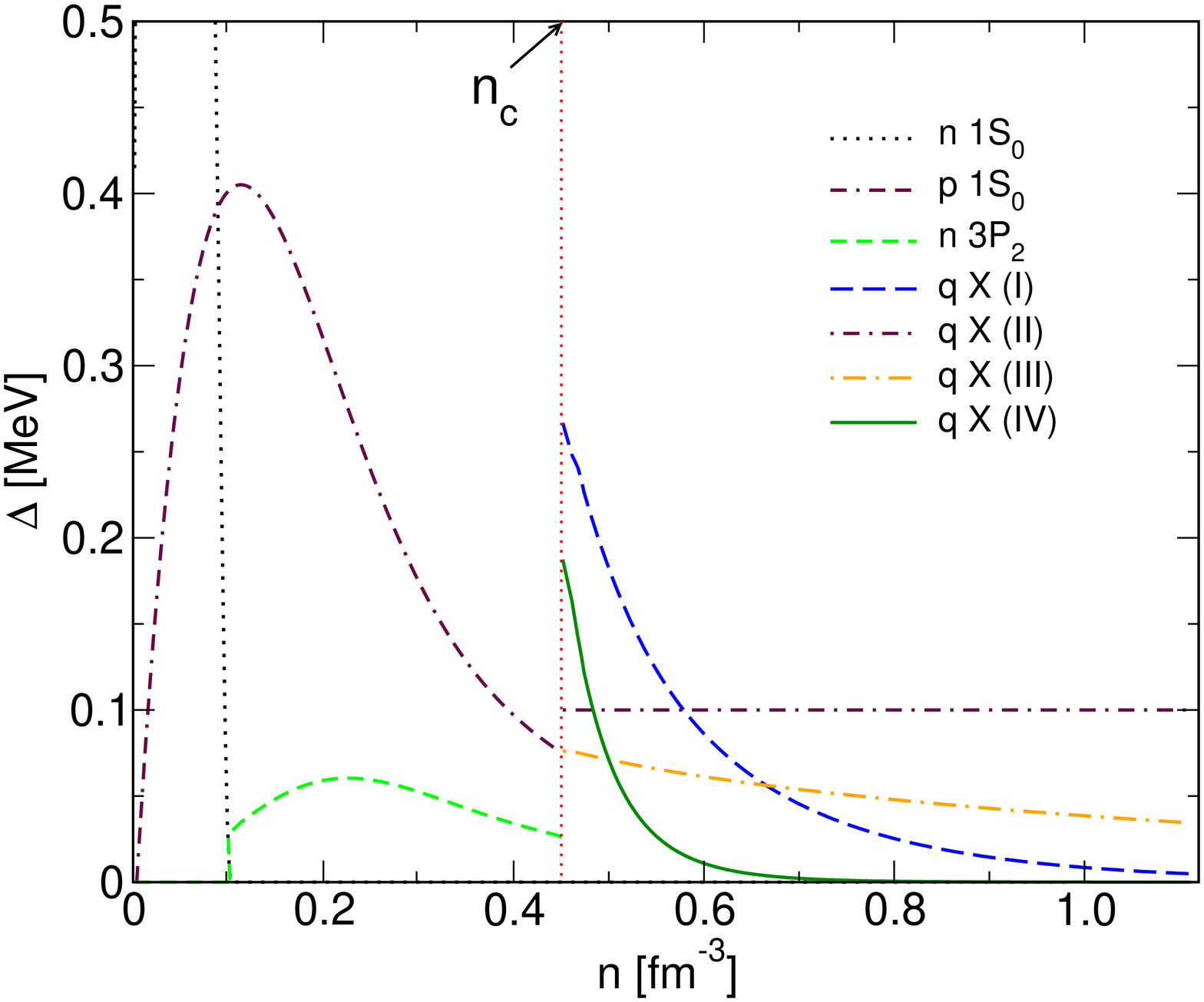} %&
\hfill 
\includegraphics[keepaspectratio,height=0.50\textwidth, angle=-90]
{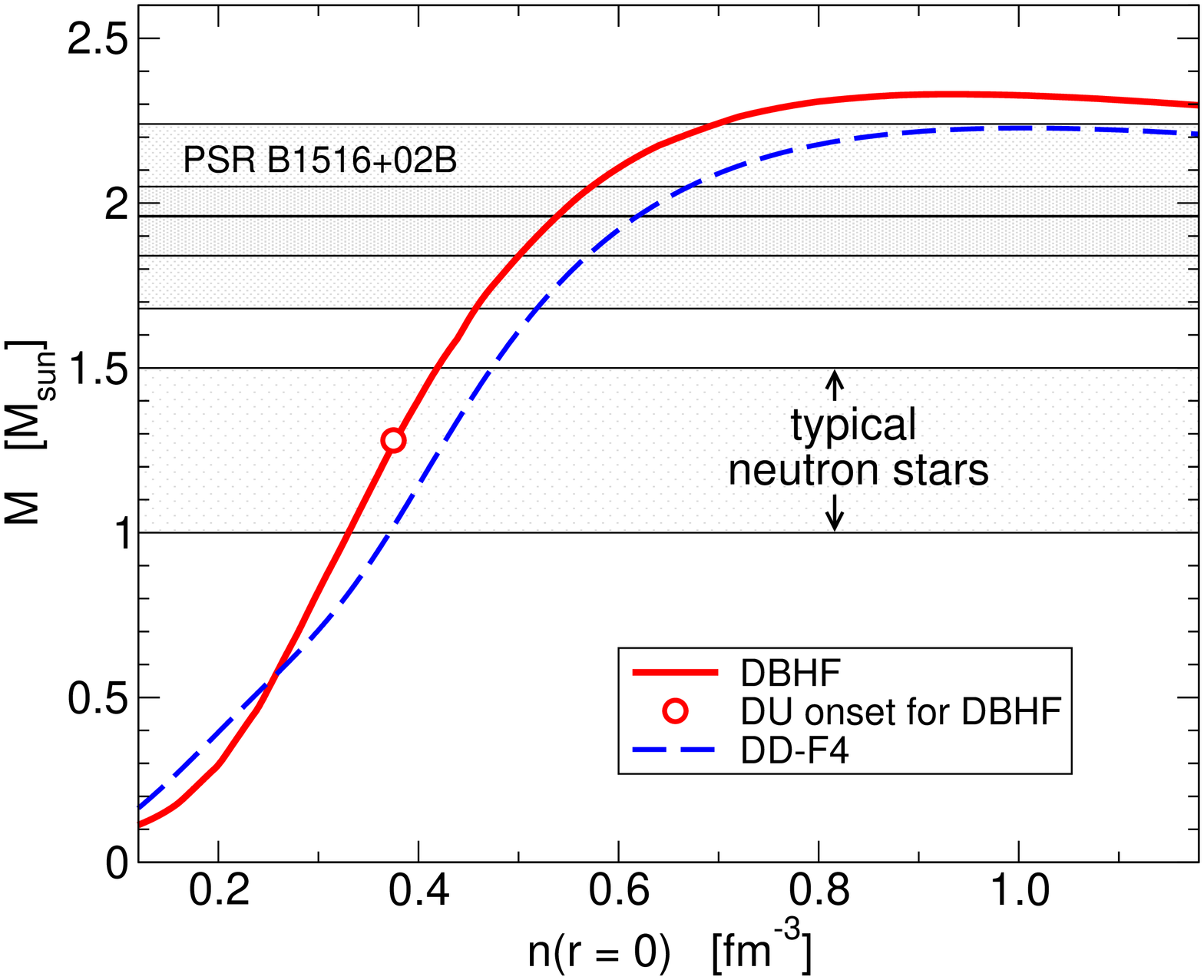} 
\caption[a]{Density dependence of the pairing gaps in nuclear matter together
  with that of the hypothetical X-gap in quark matter (left). Mass--central
  density relation for the two hadronic EoS models DBHF and DD-F4 (right). The
  dot indicates the onset of the DU process.}
\label{eos-blaschke-fig:gaps}
\end{figure}

The possibilities of pion condensation and of other so called exotic processes
are included in the calculations for purely hadronic stars but do not occur in
the hybrid star calculations since the critical density for pion condensation
exceeds that for deconfinement in our case.\cite{Blaschke:2004vq} While the
hadronic DU process occurs in the DBHF model EoS for all neutron stars with
masses above $1.27~M_\odot$, it is not present at all in the DD-F4 model, see
the right panel of Fig.\ \ref{eos-blaschke-fig:gaps}. We account for the
specific heat and the heat conductivity of all particle species whose presence
is predicted by $\beta$-equilibrium. In addition, in the case of quark matter
the contributions of massless and massive gluon-photon modes are taken into
account too.

In the 2SC phase only the contributions of quarks forming Cooper pairs (say
red and green) are suppressed via huge diquark gaps, while those of the
remaining unpaired blue color lead to a so fast cooling that the hybrid
cooling scenario becomes unfavorable.\cite{Grigorian:2004jq} We thus
assume the existence of a weak pairing channel such that in the dispersion
relation of hitherto unpaired blue quarks a small residual gap can appear. We
call
this gap $\Delta_X$ and show that for a successful description of the cooling
scenario $\Delta_X$ has to have a density dependence. We have studied the
ansatz $\Delta_{\mathrm{X}}= \Delta_0 \, \exp{\left[-\alpha\, (\mu/ \mu_c -
    1)\right]} $, where $\mu$ is the quark chemical potential, $\mu_c=330$
MeV. For the analyses of possible models we vary the values of $\alpha$ and
$\Delta_0$, given in the Table 1 of Ref.\ \refcite{Popov:2005xa} and shown in
the left panel of Fig.\ \ref{eos-blaschke-fig:gaps}.
 
\section{Temperature-Age (TA) Test} 

We consider the cooling evolution of young neutron stars with ages $t \sim
10^3 - 10^6$~yr which is governed by the emission of neutrinos from the
interior for $t \lsimB 10^5$ yr and thermal photon emission for $t \gsimB
10^5$ yr. The internal temperature is on the order of $T \sim 1$ keV. This is
much smaller than the neutrino opacity temperature $T_{\rm opac} \sim 1$ MeV
as well as the critical temperatures for superconductivity in nuclear ($T_c
\sim 1$ MeV) or quark matter ($T_c \sim 1 - 100$ MeV). Therefore, the
neutrinos are not trapped and the matter is in a superconducting state. In
Fig.\ \ref{eos-blaschke-fig:gaps} we show the density dependence of the
pairing gaps in nuclear matter\cite{Takatsuka:2004zq,Blaschke:2004vq} together
with those of the hypothetical X-gap in quark
matter.\cite{Blaschke:2004vr,Grigorian:2004jq,Popov:2005xa} The phase
transition occurs at the critical density $n_c = 2.75~n_0=0.44$ fm$^{-3}$.

\begin{figure}[tb]  
\centering 
\includegraphics[keepaspectratio,height=0.49\textwidth,
angle=-90]{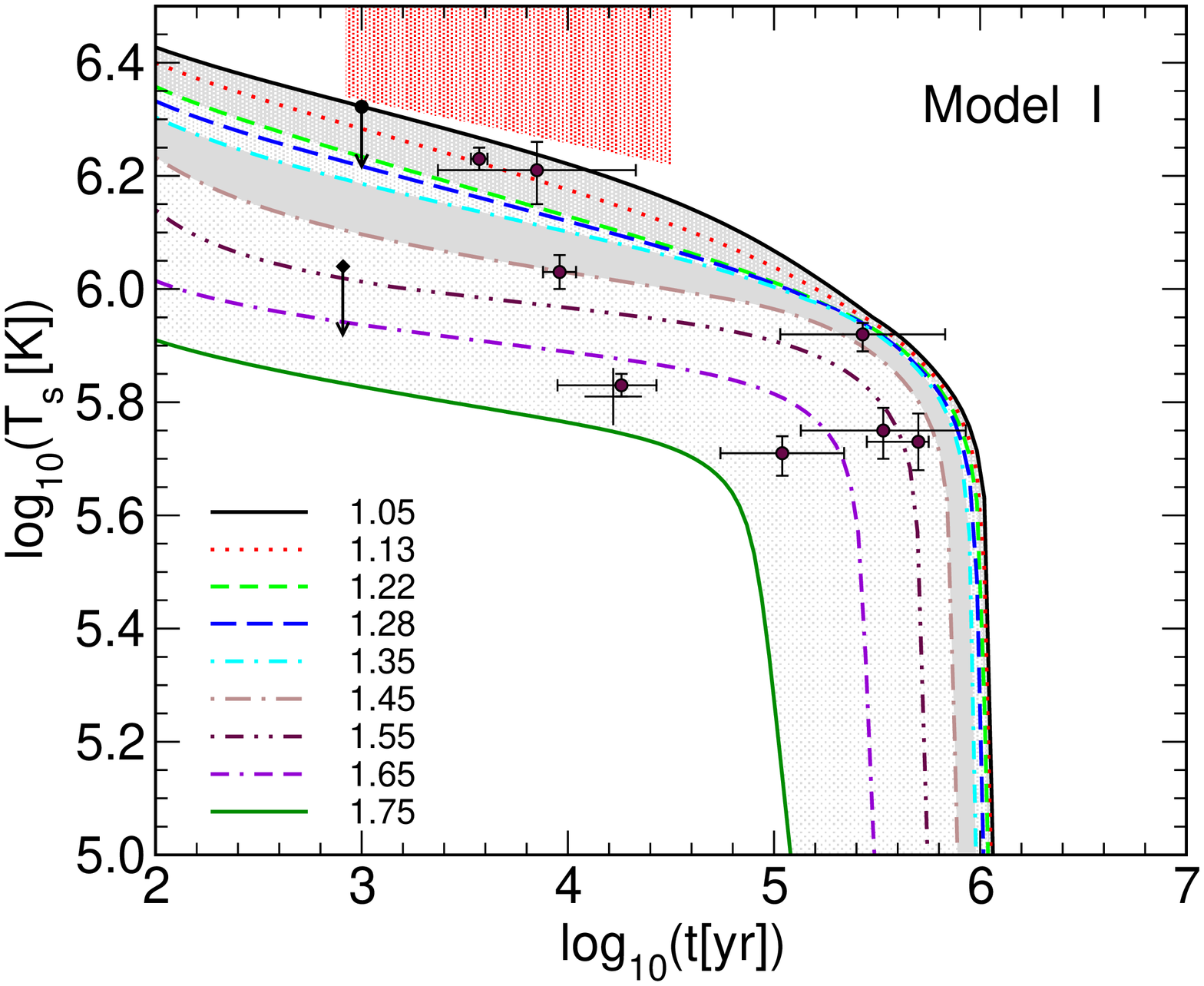} %&
\hfill
\includegraphics[keepaspectratio,height=0.49\textwidth,
angle=-90]{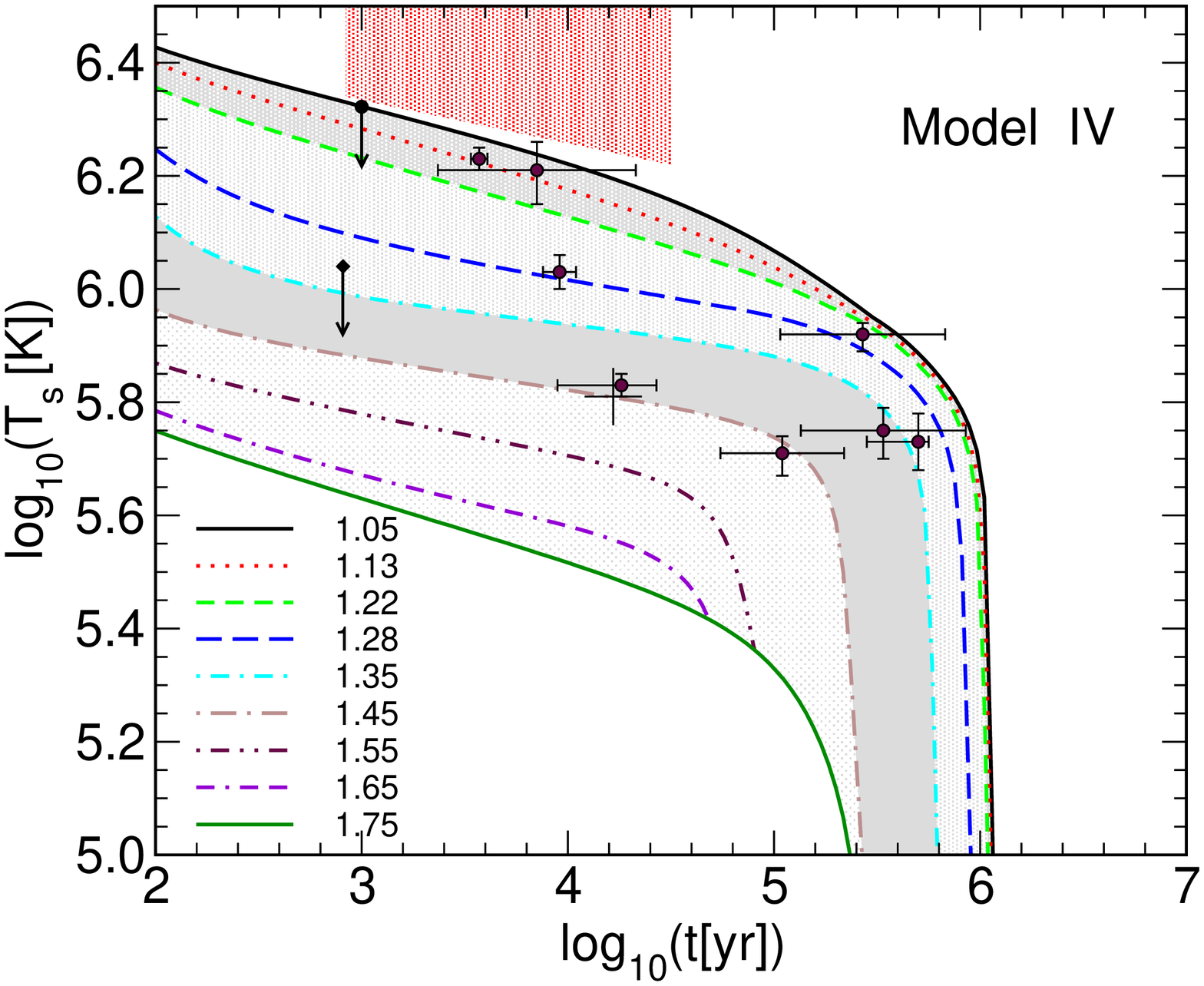}
\caption[a]{Cooling curves for hybrid star configurations with 2SC+X
  pairing pattern and X-gap model I (left) versus model IV
  (right). For the gaps see the left panel of Fig.\
  \ref{eos-blaschke-fig:hc1}.  The grey value for the shading of the
  mass bin areas corresponds to the probability for that mass bin
  value in the population synthesis model of Ref.\ \refcite{Popov:2004ey},
  for details see Ref. \refcite{Blaschke:2006gd}.
  \label{eos-blaschke-fig:qc1}}
\end{figure}

In Fig.\ \ref{eos-blaschke-fig:hc1} we present temperature--age (TA) diagrams
for two different hadronic models. Figure \ref{eos-blaschke-fig:qc1} shows the
TA diagram for two hybrid star cooling models which are presented in
Ref. \refcite{Popov:2005xa}. We note that the $T_m-T_s$ relationship between
the temperatures of the inner crust and the stellar surface has been chosen
according to a Tsuruta's formula, for details see \cite{Blaschke:2004vq}.  The
``TA test'' is fulfilled when each data point ought to be explained with a
cooling curve of an admissible configuration.  The TA data points are taken
from Ref.\ \refcite{Page:2004fy}. The hatched trapeze-like region represents
the brightness constraint (BC)\cite{Grigorian:2005fd}. For each model nine
cooling curves are shown for configurations with mass values corresponding to
the binning of the population synthesis calculations explained in Ref.\
\refcite{Popov:2005xa}.
\begin{figure}[tb]
\centering
\includegraphics[keepaspectratio,height=0.49\textwidth,
angle=-90]{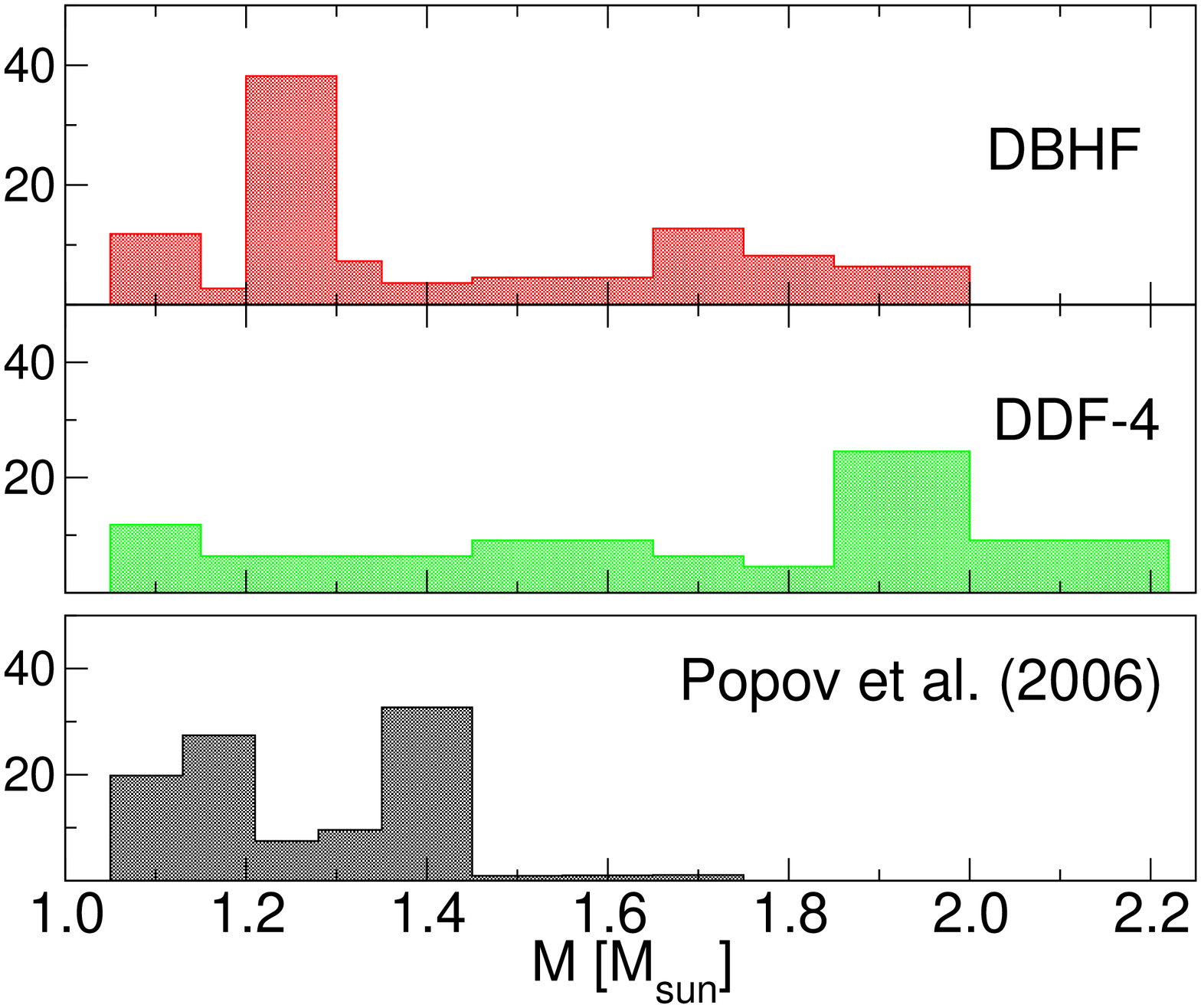}\hfill
\includegraphics[keepaspectratio,height=0.49\textwidth,
angle=-90]{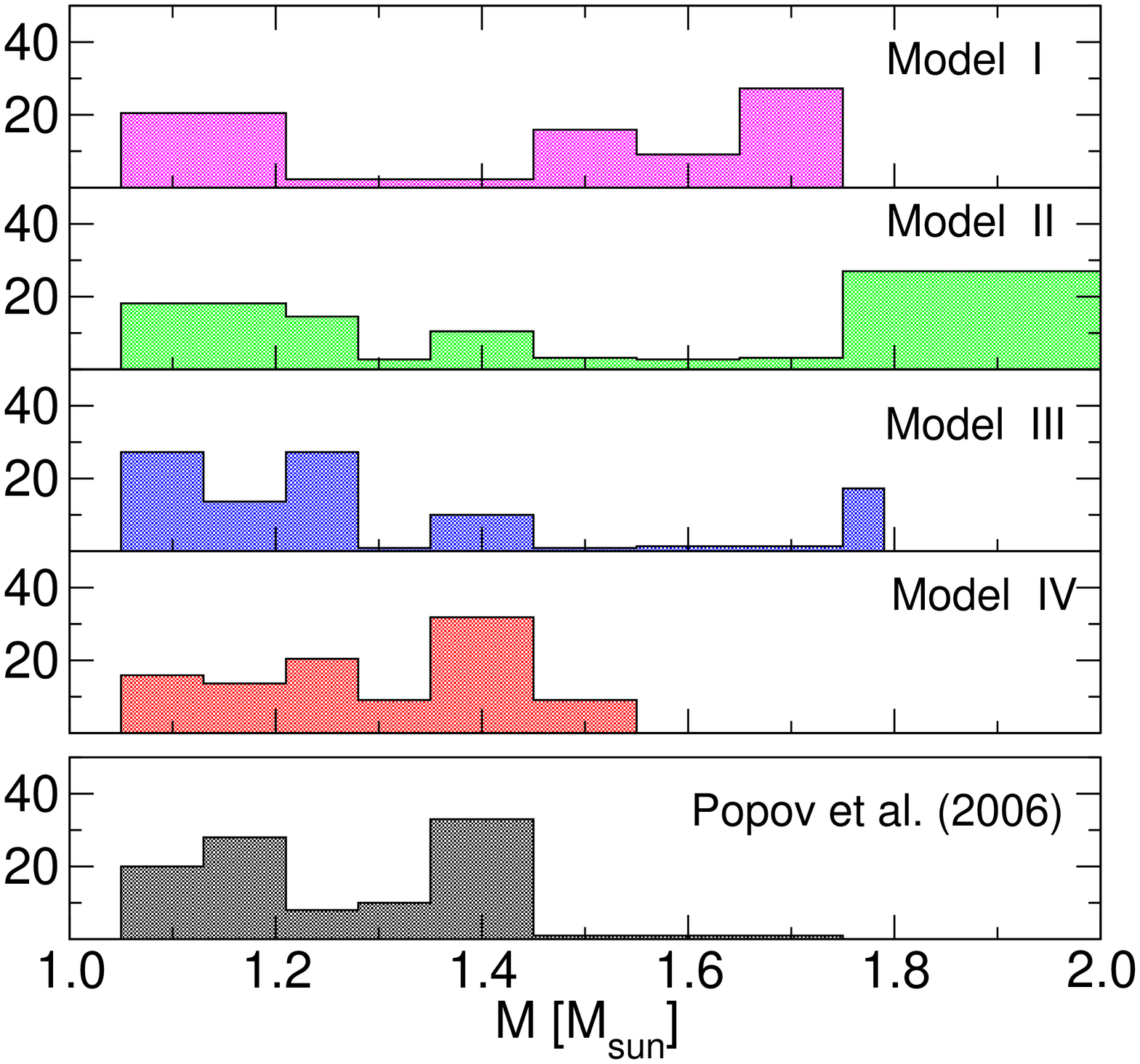}
\caption[a]{NS mass spectra extracted from the distribution
of cooling data for both hadronic EoS models (left panel) and for
hybrid stars with X-gap models I-IV (right panel). For comparison,
the mass distribution of young, nearby NS from the population
synthesis of Popov et al. \cite{Popov:2005xa} is shown at the
bottom of the panels. 
\label{fig:masshb}}
\end{figure}
A so called logN--LogS distribution constraint has been considered in Ref.\
\refcite{Popov:2005xa} for the hybrid cooling scenario above, and it has been
suggested to use the marking of TA diagram with five grey values in order to
encode the likelihood that stars in that mass interval can be found in the
solar neighborhood, in accordance with the population synthesis scenario, see
Fig.\ \ref{eos-blaschke-fig:qc1}.  The darkest grey value, for example,
corresponds to the most populated mass interval $1.35$ to $ 1.45~M_\odot$
predicted by the mass spectrum used in population synthesis.  It has been
suggested in Ref.\refcite{Blaschke:2006gd} that a mass spectrum can be derived
from the sequences of TA curves of a given compact star cooling theory, see
Fig.~\ref{fig:masshb}.  This example demonstrates how in principle cooling
simulations together with data from observation and population synthesis
simulations could discriminate pairing patterns for quark matter phases.  A
recent review\cite{Alford:2006vz} gives a flavor of the fascinating topics in
the discussion of color superconducting phases in compact stars.  Many new
developments in the theory and observations of compact stars like, e.g., the
discussion of single flavor\cite{Blaschke:2008br} and strange quark
matter\cite{Page:2005ky} phases for providing possible deep crustal heating
mechanisms to explain the puzzling phenomena of superbursts and cooling of
X-ray transients\cite{Stejner:2006tj,Shternin:2007md} will deepen our
understanding of the high-density nuclear EoS.

\section*{Acknowledgments}

We thank our collaborators, in particular H.\ Grigorian, G.\ Poghosyan, S.\
Popov, F. Sandin, and D.\ Voskresensky for their comments and for their
contributions to the results reported here.  D.\ Blaschke was supported in
part by the Polish Ministry for Science and Higher Education under contract
No.~N~N202~0953~33. T.\ Kl{\"{a}}hn is grateful for partial support from the
Department of Energy, Office of Nuclear Physics, contract no.\
DE-AC02-06CH11357.  The research of F.\ Weber is supported by the National
Science Foundation under Grant PHY-0457329, and by the Research Corporation.

%\section*{References}
%\bibliography{biblio}
%\bibliographystyle{unsrt}

\end{document}